\documentclass[twoside,journal,twocolumn,final]{IEEEtran}
\usepackage[T1]{fontenc}
\usepackage[latin1]{inputenc}
\usepackage{graphicx}


\begin{document}

\markboth{IEEE Transactions on Audio, Speech \& Language Processing, Vol. 15,
No. 3, March 2007}{Valin: On Adjusting the Learning Rate in Frequency Domain Echo Cancellation
With Double-Talk}

\title{On Adjusting the Learning Rate in Frequency Domain Echo Cancellation
With Double-Talk}

\author{Jean-Marc Valin, \IEEEmembership{Member, IEEE}
\thanks{The author is with the CSIRO ICT Centre and Xiph.org Foundation (email: jmvalin@jmvalin.ca)}
\thanks{\copyright 2007 IEEE.  Personal use of this material is permitted. Permission from IEEE must be obtained for all other uses, in any current or future media, including reprinting/republishing this material for advertising or promotional purposes, creating new collective works, for resale or redistribution to servers or lists, or reuse of any copyrighted component of this work in other works.}
}

\maketitle
\setcounter{page}{1030}
\begin{abstract}
One of the main difficulties in echo cancellation is the fact that
the learning rate needs to vary according to conditions such as double-talk
and echo path change. In this paper we propose a new method of varying
the learning rate of a frequency-domain echo canceller. This method
is based on the derivation of the optimal learning rate of the NLMS
algorithm in the presence of noise. The method is evaluated in conjunction
with the multidelay block frequency domain (MDF) adaptive filter.
We demonstrate that it performs better than current double-talk detection
techniques and is simple to implement. 
\end{abstract}

\section{Introduction}

\PARstart{R}{obust} echo cancellation requires a method for adjusting
the learning rate to account for the presence of noise and/or interference
in the signal. Most echo cancellation algorithms attempt to detect
double-talk conditions and then react by freezing the adaptation of
the adaptive filter. 

The most commonly used double-talk detection algorithm, proposed by
Gänsler \cite{Gansler1996}, is based on the coherence between the
far end and the near end signals. This algorithm has two main drawbacks.
Firstly, the detection threshold is dependent on the echo path loss,
the energy ratio between the talkers and the noise. Secondly, the
estimation of the coherence requires good knowledge (or estimation)
about the echo delay. The double-talk detector proposed by Benesty
\cite{Benesty2000} removes the need for explicit delay estimation
and generally reduces the required complexity. However, the simplification
used for computing the decision variable $\xi^{(2)}$ is based on
the assumption that the filter has converged. This assumption does
not hold when the echo path changes.

In this paper, we propose a new approach to make adaptive echo cancellation
robust to double-talk. Instead of attempting to explicitly detect
double-talk conditions, as in \cite{Gansler1996,Benesty2000}, we
use a continuous learning rate variable. The learning rate is adjusted
as a function of the interference (noise and double-talk) as well
as the misadjustment of the filter. This is done by deriving the optimal
learning rate of the normalized least mean square (NLMS) filter in
the presence of noise and applying the result to the multidelay block
frequency domain (MDF) adaptive filter \cite{Soo1990}. Although techniques
for updating the learning rate using gradient-adaptive algorithms
have also been proposed in the past \cite{Ang2001,Mandic2004}, in
this paper, we focus on designing the learning rate to react quickly
in double-talk conditions.

In Section \ref{sec:Optimal-NLMS-Adaptation}, we derive the optimal
learning rate for the NLMS algorithm in presence of noise. In Section
\ref{sec:Application-to-MDF}, we propose a technique for adjusting
the learning rate of the MDF algorithm based on the derivation obtained
for the NLMS filter. Experimental results and a discussion are presented
in Section \ref{sec:Results} and Section \ref{sec:Conclusion} concludes
this paper.

\section{Optimal NLMS learning Rate in Presence of Noise\label{sec:Optimal-NLMS-Adaptation}}

From an information theoretic point of view, we know that as long
as an adaptive filter is not perfectly adjusted, the error signal
always contains some information about the exact (time-varying) filter
weights $w_{k}(n)$. However, the amount of new information about
$w_{k}(n)$ decreases with the amount of noise in the microphone signal
$d(n)$. In the case of the NLMS filter, it means that the stochastic
gradient becomes less reliable when the noise increases or when the
filter misadjustment decreases (as the filter converges). In this
section, we derive the optimal learning rate for the general case
of the complex NLMS algorithm.

\begin{figure}
\begin{center}\includegraphics[width=1\columnwidth,keepaspectratio]{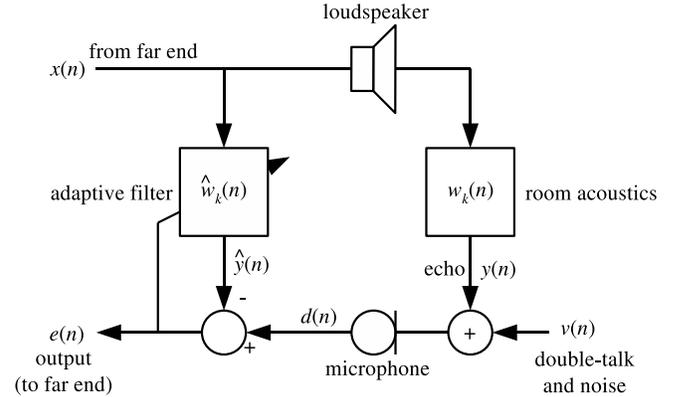}\end{center}

\caption{Block diagram of echo cancellation system\label{cap:Block-diagram}}
\end{figure}

The complex NLMS filter (Fig. \ref{cap:Block-diagram}) of length
$N$ is defined as:\begin{equation}
e(n)=d(n)-\hat{y}(n)=d(n)-\sum_{k=0}^{N-1}\hat{w}_{k}(n)x(n-k)\label{eq:NLMS_filter}\end{equation}
with adaptation step \cite{HaykinAFT}:\begin{eqnarray}
\!\!\!\!\!\!\hat{w}_{k}\!(\! n\!+\!1\!) & \!\!\!\!= & \!\!\!\!\hat{w}_{k}(n)\!+\!\mu\frac{e(n)x^{*}(n-k)}{\sum_{i=0}^{N-1}\left|x(n-i)\right|^{2}}\label{eq:NLMS_adapt1}\\
 & \!\!\!\!= & \!\!\!\!\hat{w}_{k}(n)\!+\!\mu\frac{\left(d(n)\!-\!\sum_{i}\!\!\hat{w}_{i}(n)x(n\!-\! i)\!\right)\! x^{*}(n\!-\! k)}{\sum_{i=0}^{N-1}\left|x(n-i)\right|^{2}}\label{eq:NLMS_adapt2}\end{eqnarray}
where $x(n)$ is the far-end signal, $\hat{w}_{k}(n)$ are the estimated
filter weights at time $n$ and $\mu$ is the learning rate.

Considering the error on the filter weights $\delta_{k}(n)=\hat{w}_{k}(n)-w_{k}(n)$,
and knowing $d(n)=v(n)+\sum_{k}w_{k}(n)x(n-k)$, then (\ref{eq:NLMS_adapt2})
can be re-written as:\begin{equation}
\delta_{k}(n+1)\!=\!\delta_{k}(n)+\mu\frac{\left(v(n)-\sum_{i}\!\delta_{i}(n)x(n\!-\! i)\right)x^{*}(n\!-\! k)}{\sum_{i=0}^{N-1}\left|x(n-i)\right|^{2}}\label{eq:NLMS_error}\end{equation}

At each time step, the filter misadjustment $\Lambda(n)=\sum_{k}\delta_{k}^{*}(n)\delta_{k}(n)$
can be obtained as:

\begin{equation}
\Lambda(\! n\!+\!1\!)\!=\!\sum_{k=0}^{N-1}\left|\delta_{k}(n)\!+\!\mu\frac{\!\left(\! v(n)-\sum_{i}\!\delta_{i}(n)x(\! n\!-\! i)\right)x^{*}(\! n\!-\! k)}{\sum_{i=0}^{N-1}\left|x(n-i)\right|^{2}}\!\right|^{2}\label{eq:NLMS_misadjust}\end{equation}
By making the (strong) assumption that $x(n)$ and $v(n)$ are white
noise signals uncorrelated to each other, we find that:\begin{equation}
\footnotesize{E\!\left\{ \Lambda(n+1)|\Lambda(n),\! x(n)\right\} \!=\!\Lambda(n)\!\!\left[\!1\!-\!\frac{2\mu}{N}\!+\!\frac{\mu^{2}}{N}\!+\!\frac{\mu^{2}\sigma_{v}^{2}}{\Lambda(n)\!\!\sum_{i=0}^{N-1}\!\!\left|x(n-i)\right|^{2}}\!\right]}\label{eq:Misadjust_expectation}\end{equation}
where the expectation operator $E\left\{ \cdot\right\} $ is only
taken over $v(n)$ at this point and $\sigma_{v}^{2}=E\left\{ \left|v(n)\right|^{2}\right\} $.
Because (\ref{eq:Misadjust_expectation}) is a convex function, the
expected misadjustment can be minimized with respect to $\mu$ by
solving $\partial E\left\{ \Lambda(n+1)\right\} /\partial\mu=0$ with
$\Lambda(n)\neq0$: \begin{equation}
\frac{-2}{N}+\frac{2\mu}{N}+\frac{2\mu\sigma_{v}^{2}}{\Lambda(n)\sum_{i=0}^{N-1}\left|x(n-i)\right|^{2}}=0\label{eq:Solving_optimal_rate}\end{equation}
This leads to the conditional optimal learning rate (conditioned on
the current misadjustment and the far-end signal):\begin{equation}
\mu_{opt}(n)=\frac{1}{1+\frac{\sigma_{v}^{2}}{\Lambda(n)(1/N)\sum_{i=0}^{N-1}\left|x(n-i)\right|^{2}}}\label{eq:NLMS_optimal_rate}\end{equation}
When there is no near-end noise ($\sigma_{v}^{2}$), we can see that
(\ref{eq:NLMS_optimal_rate}) simplifies to $\mu_{opt}(n)=1$, which
is consistent with \cite{Hsia1983}. Now, considering that the expectation
of $\Lambda(n)\sum_{i=0}^{N-1}\left|x(n-i)\right|^{2}/N$ over $x(n)$
is equal to the variance $\sigma_{r}^{2}(n)$ of the residual echo
$r(n)=y(n)-\hat{y}(n)$, and knowing that the output signal variance
is $\sigma_{e}^{2}(n)=\sigma_{v}^{2}(n)+\sigma_{r}^{2}(n)$, we approximate
(the approximation becomes exact as $N$ goes to infinity) the optimal
learning rate as:\begin{equation}
\mu_{opt}(n)\approx\frac{\sigma_{r}^{2}(n)}{\sigma_{e}^{2}(n)}\label{eq:Residual-to-output-rate}\end{equation}
This means that the optimal learning rate is approximately proportional
to the residual-to-error ratio. Note that $\sigma_{e}^{2}(n)$ can
easily be estimated, however the estimation of the residual echo $\sigma_{r}^{2}(n)$
is difficult and addressed in the next section. For now if we assume
we have the estimates $\hat{\sigma}_{r}^{2}(n)$ and $\hat{\sigma}_{e}^{2}(n)$
we can choose the learning rate as:

\begin{equation}
\hat{\mu}_{opt}(n)=\min\left(\frac{\hat{\sigma}_{r}^{2}(n)}{\hat{\sigma}_{e}^{2}(n)},1\right)\label{eq:uopt_const_leakage}\end{equation}
where the upper bound is the optimal rate for the noiseless case and
reflects the fact that $\sigma_{e}^{2}(n)$ is always greater than
$\sigma_{r}^{2}(n)$.

Another result that can be obtained from (\ref{eq:Misadjust_expectation})
is that the adaptation of the filter will stall ($E\left\{ \Lambda(n+1)\right\} =\Lambda(n)$)
when: \begin{equation}
\Lambda(n)\approx\frac{\sigma_{v}^{2}}{\sigma_{x}^{2}\left(\frac{2}{\mu}-1\right)}\label{eq:Adaptation_stop}\end{equation}
where $\sigma_{x}^{2}$ is the variance of the filter input (far end)
signal. Substituting the value of $\hat{\mu}_{opt}$ in (\ref{eq:uopt_const_leakage})
into (\ref{eq:Adaptation_stop}), we obtain that upon a stall in the
filter adaptation, the residual echo is:\begin{equation}
\sigma_{r}^{2}(n)\approx\min\left(\frac{1}{2}\hat{\sigma}_{r}^{2}(n),\sigma_{v}^{2}(n)\right)\label{eq:residue_limit}\end{equation}
where the first argument of the $\min\left(\cdot\right)$ is obtained
by solving \begin{equation}
\sigma_{r}^{2}(n)=\frac{\sigma_{v}^{2}(n)}{\frac{2\left(\sigma_{r}^{2}(n)+\sigma_{v}^{2}(n)\right)}{\hat{\sigma}_{r}^{2}(n)}-1}\label{eq:residue_solving}\end{equation}
The result in (\ref{eq:residue_limit}) means that the residual echo
is bounded by the background noise and by half of the estimated residual
echo, whichever is lower. For this reason, it is important not to
overestimate the residual echo by more than 3 dB, at least during
double-talk.

\section{Application to the MDF Algorithm With Background Noise and Double-Talk\label{sec:Application-to-MDF}}

The derivation in Section \ref{sec:Optimal-NLMS-Adaptation} makes
the assumption that $x(n)$ and $v(n)$ are white noise signals. While
the assumption obviously does not hold in the case of acoustic echo
cancellation of speech signals using the NLMS algorithm, we propose
to apply it to adaptive filter algorithms that operate in the frequency
domain. In this section, we concentrate on the multidelay block frequency
domain (MDF) adaptive filter \cite{Soo1990}. The adaptation used
for the MDF algorithm (and other block frequency algorithms) is similar
to applying NLMS algorithm independently for each frequency. It has
been observed that the input signals are less correlated in time (across
consecutive FFT frames) than the original time-domain signal. Also,
the learning rate $\mu(k,\ell)$ can be made frequency-dependent.
In this section, variables $\hat{Y}(k,\ell)$ and $E(k,\ell)$ are
the frequency-domain counterparts of $\hat{y}(n)$ and $e(n)$, where
$k$ is the frequency index and $\ell$ is the frame index.

Assuming the signals for each frequency of the MDF algorithm are uncorrelated
in time, we approximate the optimal frequency-dependent learning rate
by:\begin{equation}
\mu_{opt}(k,\ell)\approx\frac{\sigma_{r}^{2}(k,\ell)}{\sigma_{e}^{2}(k,\ell)}\label{eq:uopt_frequency}\end{equation}
where $k$ is the discrete frequency and $\ell$ is the frame index.
In order to estimate the residual echo $\sigma_{r}^{2}(k,\ell)$,
we make the assumption that the adaptive filter has a frequency-independent
leakage coefficient $\eta(\ell)$ that represents the misadjustment
of the filter. This leads to the estimate:\begin{equation}
\hat{\sigma}_{r}^{2}(k,\ell)=\hat{\eta}(\ell)\hat{\sigma}_{\hat{Y}}^{2}(k,\ell)\label{eq:freq_leakage}\end{equation}
where $\hat{\eta}(\ell)$ is the estimate leakage coefficient. The
advantage of this formulation is that it factors the residual estimation
$\sigma_{r}^{2}(k,\ell)$ into a slowly-evolving, but difficult to
estimate term ($\hat{\eta}(\ell)$) and a rapidly-evolving, but easy
to estimate term ($\hat{\sigma}_{\hat{Y}}^{2}(k,\ell)$). The leakage
coefficient $\eta(\ell)$ is in fact the inverse of the echo return
loss enhancement (ERLE) of the filter.

It is desirable for the learning rate to have a fast response in the
case of double-talk in order to prevent the filter from diverging
when double-talk starts. For this reason, we use the instantaneous
estimations $\hat{\sigma}_{y}(k,\ell)=\left|Y(k,\ell)\right|^{2}$
and $\hat{\sigma}_{e}(k,\ell)=\left|E(k,\ell)\right|^{2}$. Based
on (\ref{eq:uopt_const_leakage}), this leads to the learning rate:

\begin{equation}
\hat{\mu}_{opt}(k,\ell)=\min\left(\hat{\eta}(\ell)\frac{\left|\hat{Y}(k,\ell)\right|^{2}}{\left|E(k,\ell)\right|^{2}},\:\mu_{max}\right)\label{eq:uopt_estimate_freq}\end{equation}
where $\mu_{max}$ is a design parameter (always less than or equal
to 1) which puts a ceiling on the learning rate for practical purposes
and ensures that the learning rate cannot cause the adaptive filter
to become unstable.

We see from (\ref{eq:uopt_estimate_freq}) that the effects of the
filter misadjustment and the double-talk are decoupled. The learning
rate can thus react quickly to double-talk even if the estimation
of the residual echo (leakage coefficient) requires a longer time
period.

An important aspect that needs to be addressed is the initial condition.
When the filter is initialized, all the weights are set to zero, so
the $Y(k,\ell)$ signal is also zero. This causes the learning rate
computed using (\ref{eq:uopt_estimate_freq}) to be zero. In order
to start the adaptation process, the learning rate $\mu(k,\ell)$
is set to a fixed constant (we use $\mu(k,\ell)=0.25$) for a short
time equal to twice the filter length (only non-zero portions of signal
$x(n)$ are taken into account). This procedure is only necessary
when the filter is initialized and is not required in case of echo
path change.

\subsection{Leakage estimation}

We see from (\ref{eq:uopt_estimate_freq}) that the optimal learning
rate depends heavily the estimated leakage coefficient $\hat{\eta}(\ell)$.
We propose to estimate the leakage coefficient $\eta(\ell)$ by exploiting
the non-stationarity of the signals and using linear regression between
the power spectra of the estimated echo and the output signal. This
choice is based on the fact that the spectrum of the residual echo
is highly correlated with that of the estimated echo, while there
is no correlation between the spectrum of the echo and that of the
noise. 

First, a zero-mean version of the power spectra is obtained using
a first order DC rejection filter:\begin{eqnarray}
P_{Y}(k,\ell) & = & (1-\gamma)P_{Y}(k,\ell-1)\nonumber \\
 &  & +\gamma\left(\left|\hat{Y}(k,\ell)\right|^{2}-\left|\hat{Y}(k,\ell-1)\right|^{2}\right)\label{eq:zero_mean_Py}\\
P_{E}(k,\ell) & = & (1-\gamma)P_{E}(k,\ell-1)\nonumber \\
 &  & +\gamma\left(\left|E(k,\ell)\right|^{2}-\left|E(k,\ell-1)\right|^{2}\right)\label{eq:zero_mean_Pe}\end{eqnarray}
From there, $\hat{\eta}(\ell)$ is equal to the linear regression
coefficient between the estimated echo power $P_{Y}(k,\ell)$ and
output power $P_{E}(k,\ell)$:\begin{equation}
\hat{\eta}(\ell)=\frac{\sum_{k}R_{EY}(k,\ell)}{\sum_{k}R_{YY}(k,\ell)}\label{eq:leakage_estimate}\end{equation}
where the correlations $R_{EY}(k,\ell)$ and $R_{YY}(k,\ell)$ are
averaged recursively as:\begin{eqnarray}
\!\!\!\! R_{EY}(k,\ell) & \!\!\!\!\!= & \!\!\!\!\!(1\!-\!\beta(\ell))R_{EY}(k,\ell)+\beta(\ell)P_{Y}(k)P_{E}(k)\label{eq:corr_EY}\\
\!\!\!\! R_{YY}(k,\ell) & \!\!\!\!\!= & \!\!\!\!\!(1\!-\!\beta(\ell))R_{YY}(k,\ell)+\beta(\ell)\left(P_{Y}(k)\right)^{2}\label{eq:corr_YY}\\
\!\!\!\!\beta(\ell) & \!\!\!\!\!= & \!\!\!\!\!\beta_{0}\min\left(\frac{\hat{\sigma}_{\hat{Y}}^{2}(\ell)}{\hat{\sigma}_{e}^{2}(\ell)},1\right)\label{eq:corr_beta}\end{eqnarray}
$\beta_{o}$ is the base learning rate for the leakage estimate and
$\hat{\sigma}_{\hat{Y}}^{2}(\ell)$ and $\hat{\sigma}_{\hat{e}}^{2}(\ell)$
are respectively the total power of the estimated echo and the output
signal. The variable averaging parameter $\beta(\ell)$ prevents the
estimate from being adapted when no echo is present.

\subsection{Double-talk, background noise and echo path change}

It can be seen that the adaptive learning rate described above is
able to deal with both double-talk and echo path change without explicit
modelling. From (\ref{eq:uopt_estimate_freq}) we can see that when
double-talk occurs, the denominator $\left|E(k,\ell)\right|^{2}$
rapidly increases, causing an instantaneous decrease in the learning
rate that lasts only as long as the double-talk period lasts. In the
case of background noise, the learning rate depends on both the presence
of an echo signal as well as the leakage estimate. As the filter misadjustment
becomes smaller, the learning rate will also become smaller. 

One major difficulty involved in double-talk detection is the need
to distinguish between double-talk and echo path change, both of which
causing a sudden increase in the filter error signal. This distinction
is done by the leakage estimate. In conditions of double-talk, there
is little correlation between the power spectrum of the error and
that of the estimated echo, so $\hat{\eta}(\ell)$ remains small and
so does the learning rate. On the other hand, when the echo path changes,
there is a large correlation between the power spectra, which leads
to a rapid increase of $\hat{\eta}(\ell)$ that can quickly bring
the learning rate close to unity if the change is large and there
is no double-talk.

\section{Results And Discussion\label{sec:Results}}

The proposed system is evaluated in an acoustic echo cancellation
context with background noise, double-talk and a change in the echo
path. The two different impulse responses used are 1024-sample long
and measured from real recordings in a small office with both the
microphone and the loudspeaker resting on a desk. 

The proposed algorithm%
\footnote{The full source code for the proposed algorithm can be obtained as
part of the Speex software package (version 1.1.12 or later) at http://www.speex.org/%
} is compared to the Gänsler double-talk detector \cite{Gansler1996},
to the normalized cross-correlation method \cite{Benesty2000} and
to a baseline with no double-talk detection (no DTD). In the implementation
of the Gänsler algorithm, the delay estimation is performed off-line
and the coherence threshold is set to 0.3, since that value was found
to be optimal for the present case. We would expect the performance
of the Gänsler algorithm to degrade if automatic estimation of these
parameters were to be used. The optimal threshold found for the normalization
algorithm was also 0.3. It was found that choosing $\mu_{max}=0.5$
as the upper bound on the learning rate gave good results for our
algorithm. In practice, finding $\mu_{max}$ is not hard, since the
algorithm is not very sensitive to that parameter. For the other algorithms
tested, best results were achieved using $\mu=0.2$ as the learning
rate.

\begin{figure}
\begin{center}\includegraphics[width=1\columnwidth,keepaspectratio]{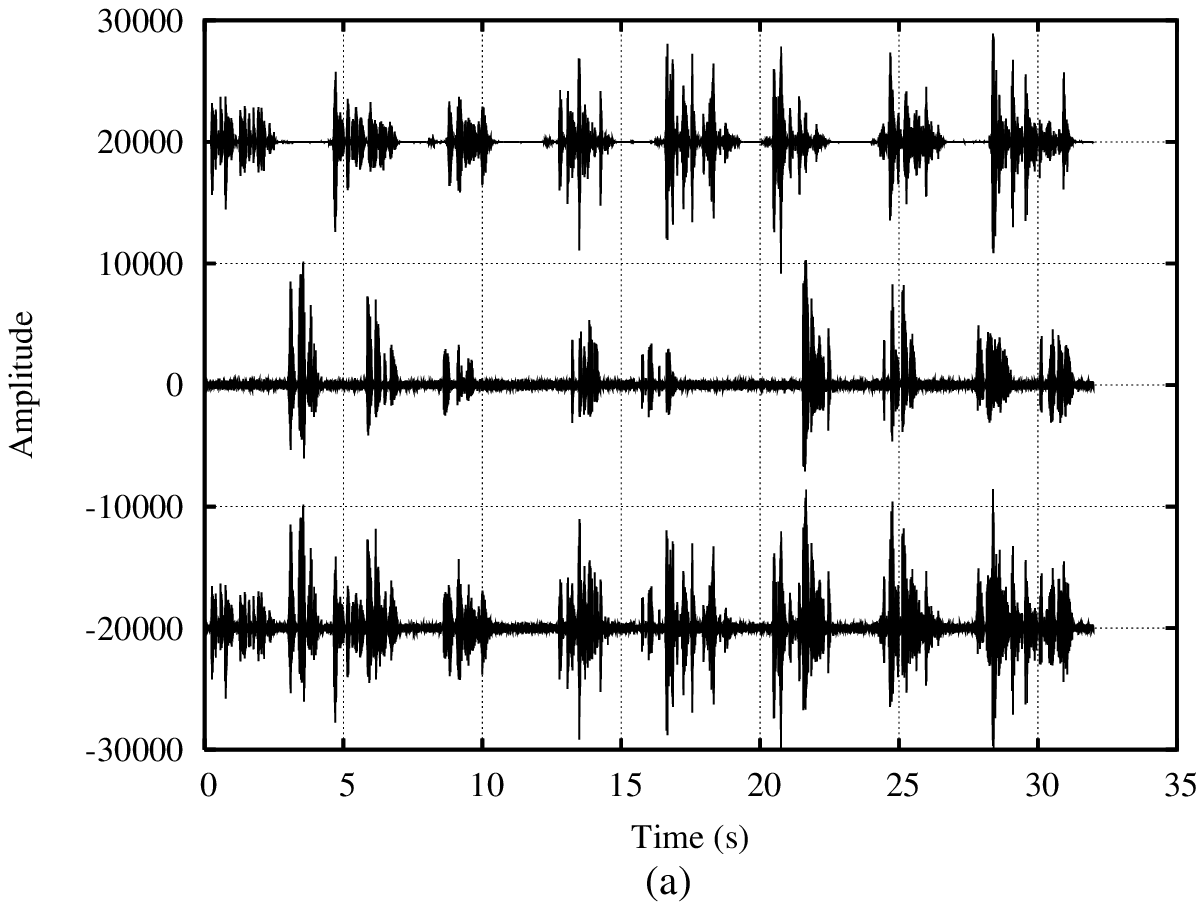}\end{center}

\begin{center}\includegraphics[width=1\columnwidth,keepaspectratio]{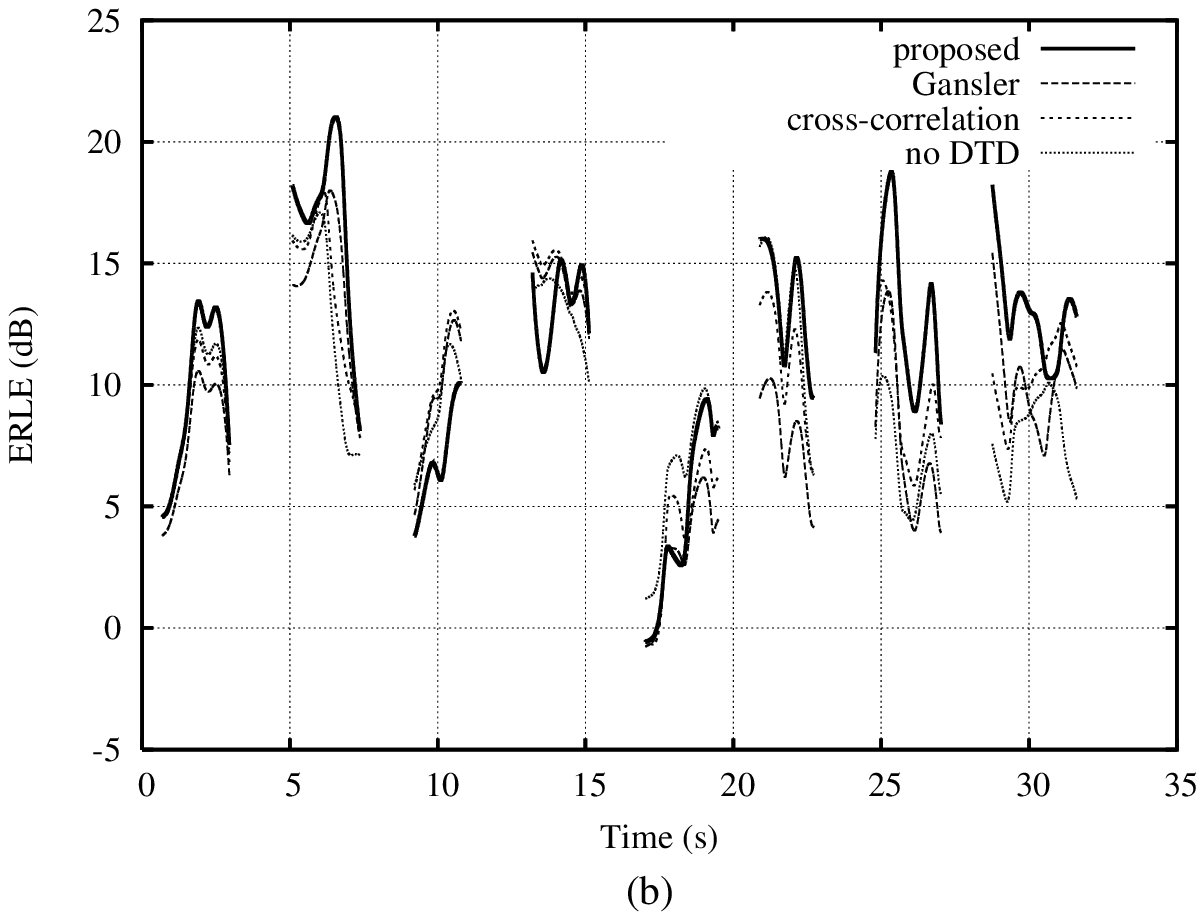}\end{center}

\caption{(a) Convolved far end signal $y(n)$ (top), near end signal $v(n)$
(middle), captured signal $d(n)$ (bottom). (b) Short-term ERLE as
a function of time. Gaps in the curves are due to the fact that the
ERLE is undefined when the far-end signal is zero.\label{cap:signals_and_ERLE}}
\end{figure}

For a typical 32-second scenario, the signals for the near-end and
far-end are shown in Fig. \ref{cap:signals_and_ERLE}a), with The
echo path changing after 16 seconds. The measured echo return loss
enhancement (ERLE) is shown in Fig. \ref{cap:signals_and_ERLE}b)
for all algorithms. Because of natural variations in the behavior
of the algorithms, it is not immediately possible to determine the
most accurate algorithm from this plot. However, we show it here to
demonstrate the behavior of our algorithm. For example, it can be
observed that when the echo path changes after 16 seconds, the proposed
algorithm re-adapts faster than the other algorithms with double-talk
detection and almost as fast as the echo canceller without double-talk
detection. 

\begin{figure}
\begin{center}\includegraphics[width=1\columnwidth,keepaspectratio]{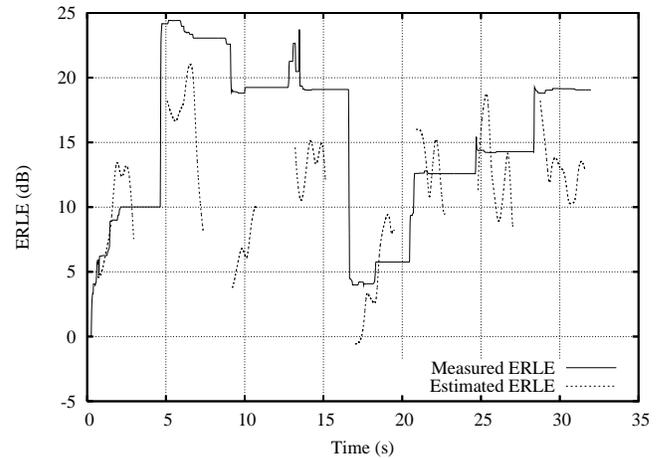}\end{center}

\caption{Echo Return Loss Enhancement (ERLE) estimate, computed as the inverse
of estimated leakage coefficient ($1/\hat{\eta}(\ell)$), compared
to the measured ERLE.\label{cap:ERLE_estim}}
\end{figure}

\begin{figure}
\begin{center}\includegraphics[width=1\columnwidth,keepaspectratio]{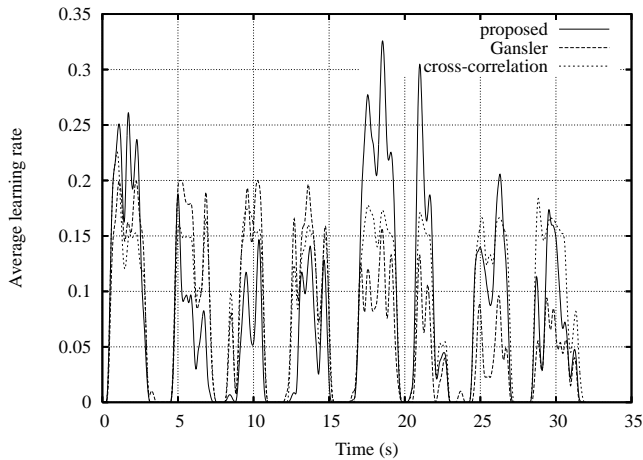}\end{center}

\caption{Learning rate averaged over a 600 ms moving window for all three
algorithms. For the proposed algorithm, the learning rate is also
averaged over all frequencies. The learning rate for the Gänsler and
normalized cross-correlation algorithms appear to to have a continuous
scale only due to the time averaging. The learning rate sometimes
goes to zero when there is no far-end signal.\label{fig:Learning-rates}}
\end{figure}

The estimate of the ERLE (computed as $1/\hat{\eta}(\ell)$) is provided
in Fig. \ref{cap:ERLE_estim}. It can be observed that the estimate
roughly follows the measured ERLE, although the estimation is obviously
noisy. Most importantly, it almost never overestimates the residual
echo (underestimate ERLE) by more than 3 dB, as is required by (\ref{eq:residue_limit}).
Also, when the echo path changes, the estimate rapidly falls toward
0~dB, which is the desired behavior. Fig. \ref{fig:Learning-rates}
shows how the learning rate varies as a function of time for all three
algorithms. The effect of the leakage estimation can be clearly observed
when the learning rate rapidly goes up after the echo path change
at $t=16\: s$, remaining well above the learning rate of the other
algorithms for about 5 seconds. It is also observed that the learning
rate goes down as the filter becomes better adapted. This is an advantage
over the Gänsler and normalized cross-correlation algorithms that
do not take into account the filter misadjustment.

\begin{figure}
\begin{center}\includegraphics[width=1\columnwidth,keepaspectratio]{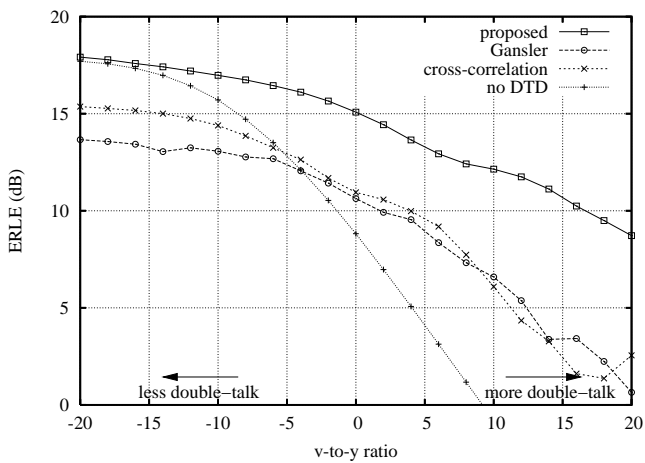}\end{center}

\caption{Steady-state ERLE (first two seconds of adaptation not considered)
as a function of the \emph{v}-to-\emph{y} signal ratio ($20\log_{10}(\sigma_{v}^{2}/\sigma_{y}^{2})$).
The filter fails to converge in the {}``no DTD'' case when the ratio
is equal or above 10 dB.\label{cap:ERLE-NFR}}
\end{figure}

\begin{figure}
\begin{center}\includegraphics[width=1\columnwidth,keepaspectratio]{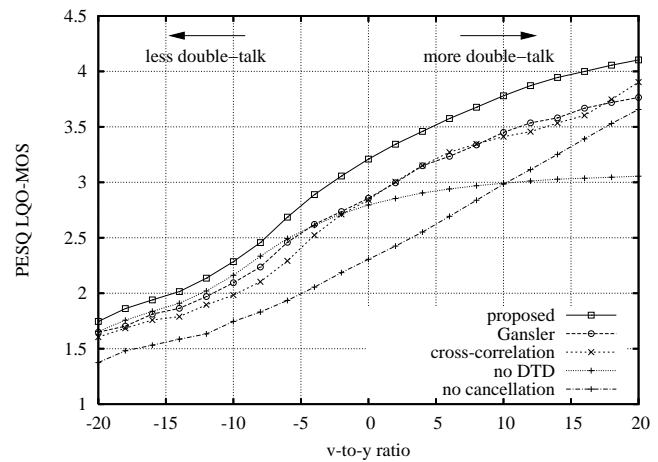}\end{center}

\caption{PESQ objective listening quality measure (LQO-MOS) as a function
of the \emph{v}-to-\emph{y} signal ratio ($20\log_{10}(\sigma_{v}^{2}/\sigma_{y}^{2})$).
The filter fails to improve in the {}``no DTD'' case when the ratio
is equal or above 10 dB.\label{cap:PESQ-NFR}}
\end{figure}

Fig. \ref{cap:ERLE-NFR} shows the average steady-state (the first
2 seconds of adaptation are not considered) ERLE for the data of Fig.
\ref{cap:signals_and_ERLE} with different ratios of near-end signal
and echo. Clearly, the proposed algorithm performs better than both
the Gänsler and normalized cross-correlation algorithms in all cases,
with an average improvement of more than 4 dB in both cases. The perceptual
quality of the output speech signal is also evaluated by comparing
it to the near field signal $v(n)$ using the Perceptual Evaluation
of Speech Quality (PESQ) ITU-T recommendations P.862 \cite{P.862}
and P.862.1 \cite{P.862.1}. The perceptual quality of the speech
shown in Fig. \ref{cap:PESQ-NFR} is evaluated based on the entire
file, including the adaptation time. It is again clear that the proposed
algorithm performs better than the reference double-talk detectors.
It is worth noting that the reason why the results in Fig. \ref{cap:ERLE-NFR}
improve with double-talk (unlike in Fig. \ref{cap:ERLE-NFR}) is that
the signal of interest is the double-talk $v(n)$, so the higher the
double-talk the less (relative) echo in the input signal.

\section{Conclusion\label{sec:Conclusion}}

We have demonstrated a novel method for adjusting the learning rate
of frequency-domain adaptive filters based on the current misadjustment
and the amount of noise and double-talk present. The proposed method
performs better than a coherence-based double-talk detector, does
not use a hard detection threshold and does not require explicit estimation
of the echo path delay. While the demonstration is done using the
MDF algorithm, we believe the technique is general enough and applicable
to other frequency-domain adaptive filtering algorithms.

In future work, the residual echo estimation in (\ref{eq:freq_leakage})
could be evaluated as a residual echo estimator for further suppression
of the echo, as proposed in \cite{Gustafsson2002}. Also, more accurate
methods for estimating the leakage coefficient should be researched.

\bibliographystyle{ieeetr}
\bibliography{echo}

\begin{thebibliography}{10}

\bibitem{Gansler1996}
T.~G{\"{a}}nsler, M.~Hansson, C.-J. Ivarsson, and G.~Salomonsson, ``A
  double-talk detector based on coherence,'' {\em IEEE Transactions on
  Communications}, vol.~44, no.~11, pp.~1421--1427, 1996.

\bibitem{Benesty2000}
J.~Benesty, D.~Morgan, and J.~Cho, ``A new class of doubletalk detectors based
  on cross-correlation,'' {\em IEEE Transactions on Speech and Audio
  Processing}, vol.~8, no.~2, pp.~168--172, 2000.

\bibitem{Soo1990}
J.-S. Soo and K.~Pang, ``Multidelay block frequency domain adaptive filter,''
  {\em IEEE Transactions on Acoustics, Speech and Signal Processing}, vol.~38,
  no.~2, pp.~373--376, 1990.

\bibitem{Ang2001}
W.-P. Ang and B.~Farhang-Boroujeny, ``A new class of gradient adaptive
  step-size lms algorithms,'' {\em IEEE Transactions on Signal Processing},
  vol.~49, no.~4, pp.~805--810, 2001.

\bibitem{Mandic2004}
D.~Mandic, ``A generalized normalized gradient descent algorithm,'' {\em IEEE
  Signal Processing Letters}, vol.~11, no.~2, pp.~115--118, 2004.

\bibitem{HaykinAFT}
S.~Haykin, {\em Adaptive Filter Theory}.
\newblock Prentice Hall, 4~ed., 2002.

\bibitem{Hsia1983}
T.~Hsia, ``Convergence analysis of {LMS} and {NLMS} adaptive algorithms,'' in
  {\em Proceedings IEEE International Conference on Acoustics, Speech, and
  Signal Processing}, vol.~8, pp.~667--670, 1983.

\bibitem{P.862}
ITU-T, {\em Perceptual evaluation of speech quality (PESQ): An objective method
  for end-to-end speech quality assessment of narrow-band telephone networks
  and speech codecs}.
\newblock International Telecommunications Union, 2001.

\bibitem{P.862.1}
ITU-T, {\em Mapping function for transforming P.862 raw result scores to
  MOS-LQO}.
\newblock International Telecommunications Union, 2003.

\bibitem{Gustafsson2002}
S.~Gustafsson, R.~Martin, P.~Jax, and P.~Vary, ``A psychoacoustic approach to
  combined acoustic echo cancellation and noise reduction,'' {\em IEEE
  Transactions on Speech and Audio Processing}, vol.~10, no.~5, pp.~245--256,
  2002.

\end{thebibliography}

\begin{biography}
[{\includegraphics[width=1in,height=1.25in]{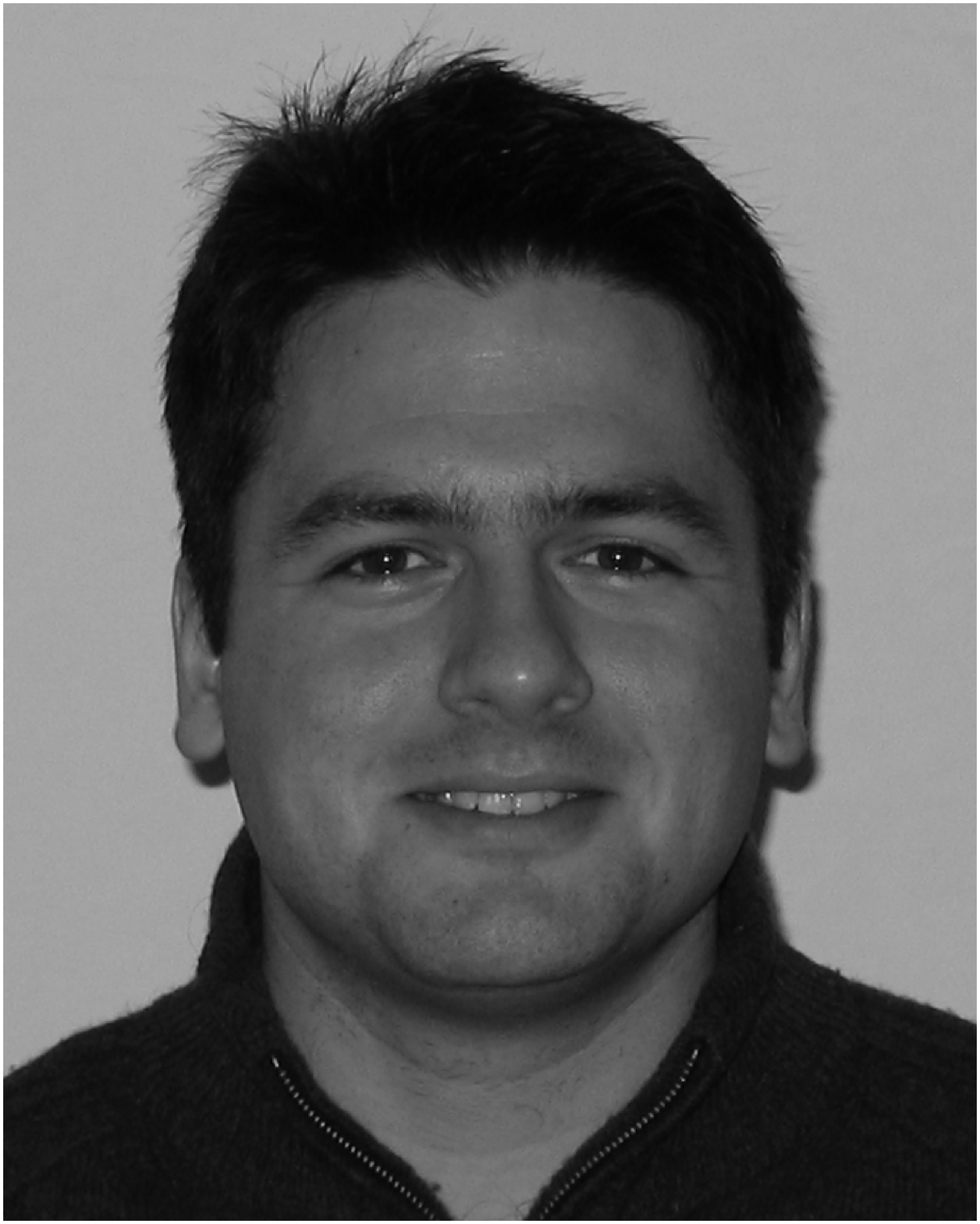}}]{Jean-Marc
Valin} (S'03-M'05) was born in Montreal, Canada in 1976. He received
his B.Eng. degree, M.Sc. degree and Ph.D. in the electrical engineering
from the University of Sherbrooke in 1999, 2001 and 2005 respectively.
His Ph.D. research focused on bringing auditory capabilities to a
mobile robotics platform, including sound source localization and
separation. Since 2005, he is a post-doctoral fellow at the CSIRO
ICT Centre in Sydney, Australia. His research topics include acoustic
echo cancellation and microphone array processing. He is the author
of the Speex speech codec. 
\end{biography}

\end{document}